\documentclass[preprint,aps]{revtex4}
\usepackage[dvips]{graphicx}
\usepackage{subfigure}
\begin{document}
\title{Classical-path integral adaptive resolution
  in molecular simulation: towards a smooth quantum-classical coupling}
\author{A.B.Poma and L.Delle Site}
\address{Max-Planck-Institute for Polymer Research,
Ackermannweg 10, D 55021 Mainz Germany.}

\begin{abstract}
Simulations that couple different classical molecular
models in an adaptive way by changing the number of degrees of freedom on the
fly, are available within reasonably consistent theoretical
frameworks. The same does not occur when it comes to classical-quantum
adaptivity. The main reason for this is the difficulty in describing a
continuous transition between the two different kind of physical principles:
probabilistic for the quantum and deterministic for the classical. Here we
report the basic principles of an algorithm that allows for a
continuous and smooth transition by employing the path integral description of
atoms.
\end{abstract}
\maketitle
\begin{figure}[H]
 \includegraphics[width=0.4\textwidth]{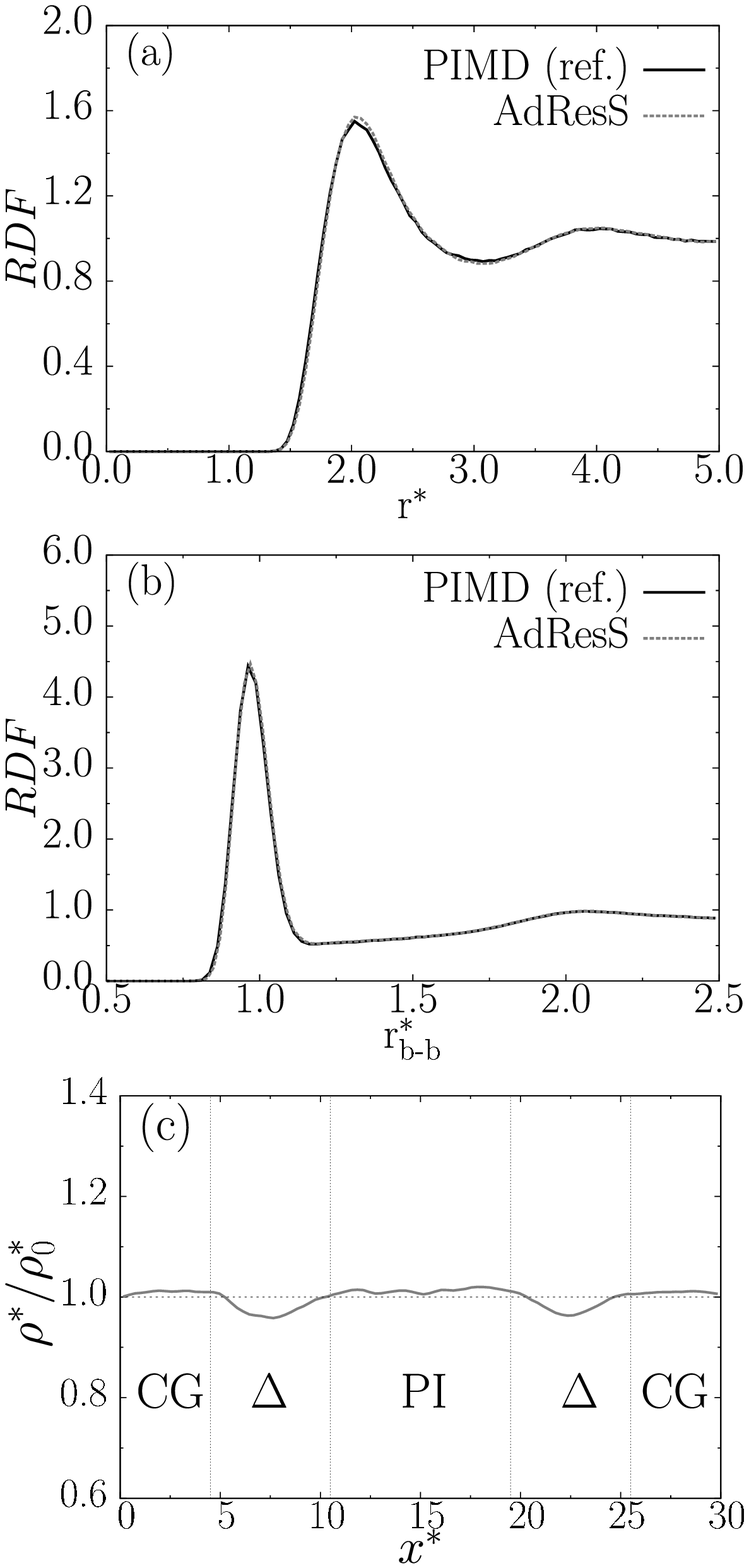}
\caption{Results for the adaptive simulation of the liquid of tetrahedral molecules at the temperature $T_{1}$. Top (a), the molecular center of mass-center of mass radial distribution function obtained with AdResS is compared with that obtained from the full path integral reference system,. Middle (b), the bead-bead radial distribution function obtained with AdResS in the quantum region compared with that of  the full path integral reference system. Bottom (c), the particle density in AdResS compared with the reference system. The density is equal to $0.1 \sigma^{-3}$ in the units reported in the Appendix.\label{t1}}
\end{figure}
\begin{figure}[H]
\includegraphics[width=0.4\textwidth]{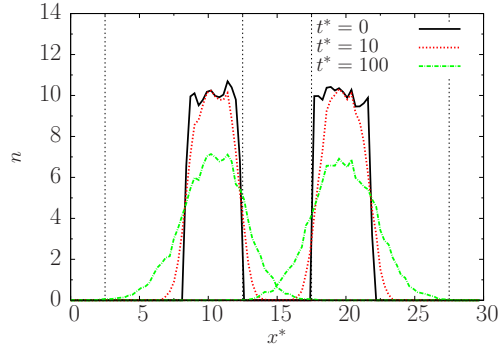}
\caption{Diffusion profile for the molecules moving from the path integral region to the coarse grained one and for molecules moving in the opposite direction, for the system at temperature $T_{1}$. The picture shows that no barriers, due to possible artifacts of the algorithm, hinter the diffusion process.\label{diff}}
\end{figure}

{\bf Introduction:} The development of adaptive resolution simulation schemes  is a subject of growing interest within the community 
of condensed matter, material and chemical physics. 
By adaptive resolution it is meant that the space is partitioned in regions
characterized by different molecular resolutions where molecules can
freely diffuse changing their representation according to the region where they are instantaneously located. In the last few years several approaches have been presented and they are characterized by different
levels of theoretical sophistication and computational complexity \cite{annurev,jcpth,ens,hyden,voth,slaz}. The
interest in this kind of approach arises from the fact that it may
efficiently tackle the problem of interplay between different scales. 
However while the adaptive process can be 
described in a reasonable way according to the basic principles of classical
dynamics and thermodynamics, the same cannot be said when quantum mechanics
enters into the game. In general, the proper coupling of quantum and classical mechanics
is known to be a non trivial (and open) problem (see e.g.\cite{giov}) and here the adaptive character adds up as a
further major difficulty \cite{nic}.
In this work we aim to develop an approach where the idea of coupling a
classical and quantum molecular model in an adaptive fashion can be made in a way that the
''probabilistic-deterministic'' discontinuity is removed and yet the adaptivity takes place in a smooth controlled numerical way.\\ 
{\bf Path Integral representation of atoms and molecules:} In this work our quantum systems are composed by
atoms described within the path integral formalism. This approach is, by
now, a standard tool in molecular dynamics and well established in
literature (see e.g. \cite{tuck}). For this reason here we do not present a technical description of the method, instead we explain why this approach is optimal to adaptively couple quantum systems with a classical bath. 
The important aspect of the path integral approach for this work is
that an atom, which is usually represented as a sphere in classical force
fields, becomes, within the path integral formalism, a classical ``polymer ring'' so that the interaction site is
delocalized into the beads of such a polymer; each bead of the polymer is
linked to its next neighbors along the chain by a harmonic potential. The elastic constant depends on the
temperature, $T$, of the system and the number of beads, $n$, used to represent each atom, $k=\frac{m(k_{B}T)^{2}}{n^{2}\hbar^{2}}$, with $k_{B}$  being , as usual, the Boltzmann constant, $m$ the mass of the atom, $\hbar$ Planck's constant.
\begin{figure}[H]
\includegraphics[width=0.4\textwidth]{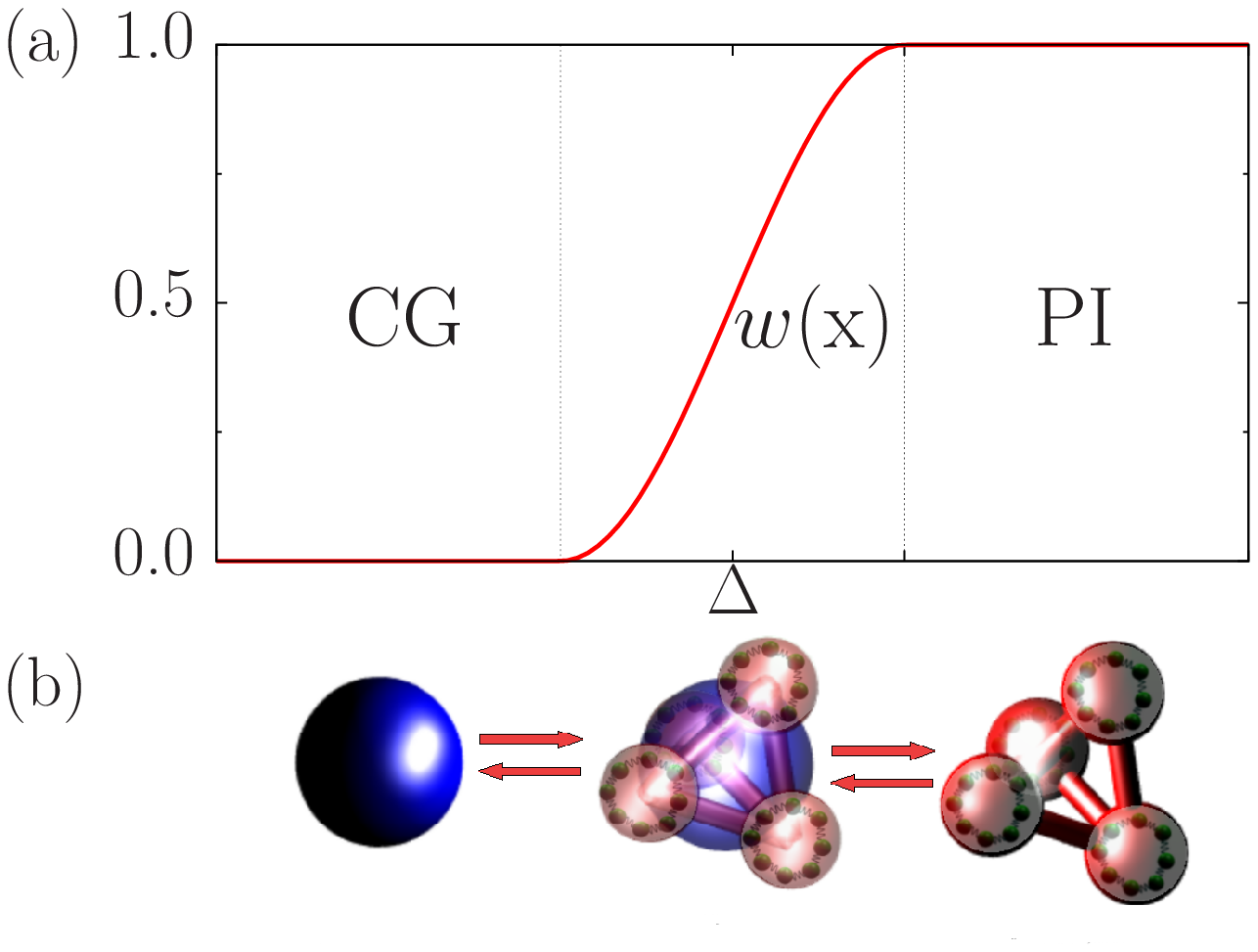}
\caption{Pictorial representation of the adaptive box and
molecular representation. The region on the left, indicated by $CG$,
is the low resolution region (coarse-grained), the central part is
the transition (hybrid) region $\Delta$, where the switching function
$w(x)$ is defined, and the region on the right,
indicated by $PI$ is the high resolution region (full path integral region).\label{cartoon1}}
\end{figure}
Such a delocalization characterizes the quantum nature of the atom. This approach has been
extensively used in molecular dynamics to account for basic quantum effects
otherwise not described by standard classical force fields \cite{berne,marx1,david2,david1,marx2,voth2}. For the adaptive process the
classical polymer representation of atoms has far reaching consequences,
because it translates the quantum-classical coupling into the coupling of two effective classical regions characterized by a
different number of (as a matter of fact) ''classical'' degrees of freedom; thus the whole machinery of classical adaptive methods would apply straightforwardly.
In this work we show that indeed this is the case. Obviously the approach here 
is suited only for a class of problems and, as it is designed now, certainly not for those problems where the quantum mechanics refers to electrons.  It must be noted that practical methods \cite{slaz,hyden2,bulo} which couple the two regimes when electrons are considered suffer from the same conceptual limitations underlined in this work and base their validity on numerical criteria only. In this specific case the conceptual discontinuity is in the arbitrary cut off of the electron wavefunction and in the non conservation of the number of electrons as the system evolves in time. 
Although we do not consider systems with electrons, the concept of mapping a quantum problem onto a classical one and then have a smooth adaptivity is anyway of general valence. In this sense this work represents a possible starting point for a general adaptive coupling of quantum and
classical description of atoms and molecules.\\
 {\bf The Adaptive scheme:} Regarding the adaptive
scheme for the classical case, the basic requirement is a controlled procedure
of changing the number of degrees of freedom based on solid physical
principles and consistent with the thermodynamic equilibrium of the overall
system \cite{pre2,jpa,premio}. In this respect the AdResS method meets the requirements above in a
(extensively tested) satisfactory way \cite{jcp,pre,jcppol,wat1,wat2}. For this reason in this work the
coupling between the polymer rings and the classical particles will be done within the AdResS framework. Below we introduce the basic features of AdResS relevant for this work.
 In the classical AdResS method the atomistic (high resolution) and the coarse-grained (low resolution) 
regime are coupled via a force interpolation (see e.g. \cite{annurev,nic}):
\begin{equation}
{\bf F}_{\alpha \beta}=w(X_\alpha)w(X_\beta){\bf
  F}_{\alpha\beta}^{atom}+[1-w(X_\alpha)w(X_\beta)]{\bf F}^{cg}_{\alpha\beta} 
\label{eqforce}
\end{equation}
where $\alpha$ and $\beta$ indicates two molecules, ${\bf F}^{atom}$ is the
force derived from the atomistic force field and  ${\bf F}^{cg}$
from the corresponding coarse-grained potential, $X$ is the $x$ coordinate of the center of mass of
the molecule and $w$ is an interpolating function which smoothly goes from $0$
to $1$ (or vice versa) in a transition region ($\Delta$) where the lower resolution is then
slowly transformed (according to $w$) in the high resolution (or vice versa),
as illustrated in Fig.\ref{cartoon1} (see also \cite{note1}).
An additional locally acting thermostat is employed to assure the overall
thermodynamic equilibrium.   
The coarse-grained potential is obtained from a reference all atom simulation
at the given thermodynamic condition via an iterative inverse Boltzmann procedure employing the molecular center of mass radial distribution \cite{florian}.
For the case of the path integral description we have exactly the same procedure as for the classical case but with the difference that instead of a classical atomistic representation of the molecule we have a molecular representation where the atoms are described as polymer rings in a path integral approach (see Fig.\ref{cartoon1}).
It follows that the coarse-grained model is derived from a full path integral 
reference simulation at the given thermodynamic condition.
Given the framework reported above the
coupling between the path integral representation and the coarse-grained model occurs via the interpolation according to $w$ between ${\bf
  F}^{pi}$ acting on the beads of the rings (which now plays the equivalent role of ${\bf F}^{atom}$ in Eq.\ref{eqforce}) and the ${\bf F}^{cg}$ derived from the coarse grained potential acting among the centers of mass of the molecules.  We tested this idea studying a liquid of tetrahedral molecules whose atomistic model was used in the original development of AdResS (see Fig.\ref{cartoon1}(b)).\\
{\bf Results of the Path integral-coarse-grained Adaptive Resolution Simulation:} We studied a system of thousand molecules at two different temperatures, indicated as $T_{1}$ and $T_{2}$. 
$T_{1}$ is the same temperature employed in the previously studied classical model and $T_{2}=\frac{T_{1}}{\sqrt{10}}$. In the Lennard-Jones units of this paper (see the Appendix or Ref.\cite{jcp}) $T_{1}=1$ and consequently $T_{2}=\frac{1}{\sqrt{10}}$. 
The different temperatures are directly related to the elastic constant of the polymers and thus, for the same number of beads per ring ($n=10$), a lower temperature means more flexibility of the polymer rings, thus we have $k_{2}=\frac{k_{1}}{10}$. For the testing purpose we have deliberately chosen $T_{2}$ because it mimics the thermodynamic conditions of a ''more quantum'' system than $T_{1}$ (see also \cite{note2}). Given the computational cost of
path integral simulations and the very extended range of tests we have used, for both systems, the number of beads per polymer, $n$, equal to $10$ \cite{note3}.
Fig.\ref{t1} reports the radial distribution functions (RDF) and the density distribution of the AdResS simulation of thousand molecules compared to a full
path integral simulation for the temperature $T_{1}$. 
They show that indeed this
coupling procedure displays the desired behaviour. In particular, in Fig.\ref{t1}(b), the comparison
between the bead-bead radial distribution function obtained with AdResS in the
quantum region and that obtained from a full path integral simulation shows
that indeed the very quantum nature of the particles in the quantum region of
AdResS is very well described and thus the ''classical bath'' of the coarse-grained molecules is indeed able to
reproduce the overall thermodynamic conditions properly. 
Regarding the density in Fig.\ref{t1}(c), the deviations with respect to the 
reference value (above all in the transition region) are the same obtained in the classical case \cite{jcp}. As for the classical AdResS case, one must be sure that indeed there is exchange of molecules between the two regimes. In fact it may occur situations where the two regimes are in equilibrium because some barriers, artifact of the method, would hinder the free exchange so that the molecules are reflected back at the transition region. Fig.\ref{diff} shows that this is not the case and the molecules diffuse between the two regions in a proper way.
\begin{figure}[H]
\includegraphics[width=0.4\textwidth]{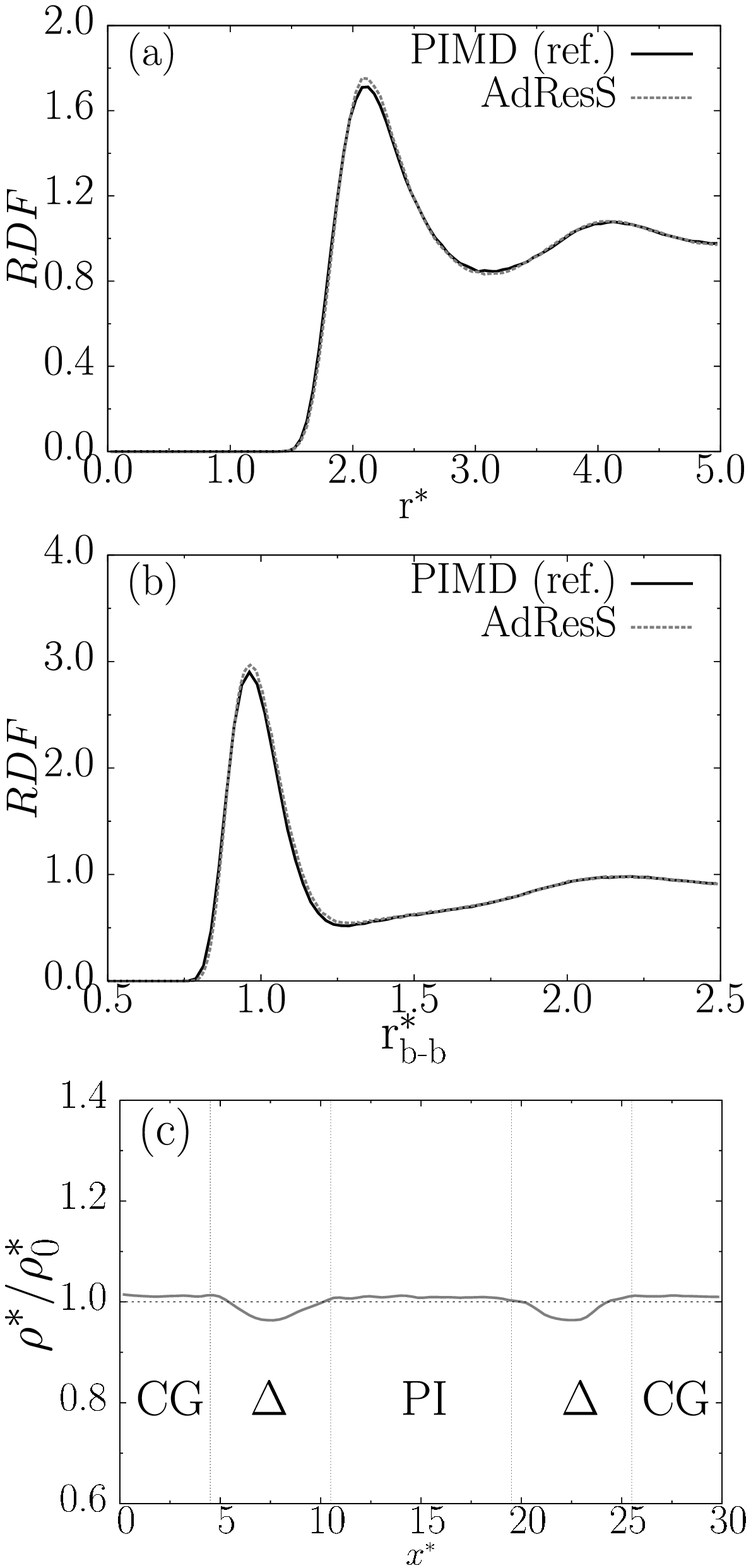}
\caption{As Fig.\ref{t1} but at the temperature $T_{2}$.\label{t2}}
\end{figure}
Fig.\ref{t2} shows the results for $T_{2}$, and also in this case the agreement with the reference full path integral simulation is satisfactory (the diffusion profile, not reported here, shows also the same behaviour as for the case $T_{1}$). The overall agreement make us confident that the principles of the adaptive algorithm can be extended to the case of path integral representation.\\
{\bf Conclusions:} In general for physical systems one should not expect the agreement found here although it
can be expected a reasonable accuracy in the same way (and with similar
limitations) of the classical AdResS.
Having shown that the idea works in a rather reasonable way, it remains to address the question of why such an approach may be useful for applications and not a mere conceptual exercise. In general, there are several problems in soft condensed matter 
where relevant properties are the results of the interconnection between the local bonding and the larger scale molecular packing. In such cases classical models may give a satisfactory description of the large scale packing but are not sufficient for the high resolution required locally and the path integral description becomes crucial \cite{marto}; this is the situation where our approach would represent an optimal tool.
Another important field of application is that involving liquid water and above all water as a solvent of large molecules \cite{strajbl}.
In particular in cases where the solute-water hydrogen bond plays a crucial role \cite{schneider, dalperaro,dana} and classical model can only partially describe the water mediated process. Moreover, combined with specific force fields \cite{voth3}, would be ideal to study local protonation/deprotonation of molecules in solution. In these cases the adaptive method would require that only the region around the solute should be treated with the path integral resolution while the rest can be treated in a coarse-grained fashion as shown here or with a hierarchy of models as already done in part within AdResS \cite{wat1,wat2,rafaw}.
Needless to say that this is only a first step towards the development of a generic scheme for quantum-classical coupling and serve as a basis for further work.\\
{\bf Appendix: The Classical atomistic model}\\
Here we briefly report the basic technical details of the force field for the tetrahedral molecules.
Each molecule is characterized by $N=4$ atoms of the same mass
$m_0$ connected by anharmonic bonds.
All atoms in the system interact according to a purely repulsive
shifted $12$-$6$ Lennard-Jones potential with a cutoff at
$2^{1/6}\sigma$:
\begin{eqnarray}
 U_{LJ}^{atom}(r_{i\alpha j\beta})=\left\{\begin{array}{rc}
4\varepsilon\bigl[\bigl(\frac{\sigma}{r_{i\alpha
      j\beta}}\bigr)^{12}-\bigl(\frac{\sigma}{r_{i\alpha j\beta}}\bigr)^6+\frac{1}{4}\bigr];
& r_{i\alpha j\beta}\le 2^{1/6}\sigma\\
                          0; & r_{i\alpha j\beta}> 2^{1/6}\sigma
                             \end{array}\right.\label{eq.8}
\end{eqnarray}
$r_{i\alpha j\beta}$ defines the distance between the atom $i\alpha$ of the molecule
$\alpha$ and the atom $j\beta$ of the molecule $\beta$. We define
$\varepsilon$ as a unit of energy. All atoms have the same excluded
volume diameter $\sigma$, where $\sigma$ is the unit of length.
Neighboring atoms of a molecule are linked via an attractive FENE potential
\begin{eqnarray}
 U_{FENE}^{atom}(r_{i\alpha j\alpha})=\left\{\begin{array}{rc}
                  -\frac{1}{2}kR_0^2\ln\bigl[1-\bigl(\frac{r_{i\alpha
 j\alpha}}{R_0}\bigl)^2\bigr]; & r_{i\alpha j\alpha}\le R_0\\
                       \infty; & r_{i\alpha j\alpha}> R_0
                  \end{array}\right.\label{eq.9}
\end{eqnarray}
$R_0=1.5\sigma$ is the divergence length  and $k=30\varepsilon/\sigma^2$ the stiffness.         

{\bf Acknowledgments}\\
We thank C.Peter, D.Donadio, M.Praprotnik and K.Kremer for discussions and for a critical reading of the manuscript; D.Manolopoulos for the path integral code and a critical reading of the manuscript.
ABP acknowledge the financial support of the DAAD, LDS that of the MMM initiative of the Max-Planck Society.


\begin{thebibliography}{9}
\bibitem{note1}
In practical terms this means that when a molecules goes from the path integral to the coarse grained region, the force acting directly on the beads will smoothly vanish and that acting on the molecular center of mass (coarse-grained force) will increase. Vice versa when a coarse grained molecule moves towards the path integral region the force acting on the beads is smoothly reactivated and that acting on the molecular center of mass smoothly vanishes. From the technical point of view, while the atoms and the corresponding polymer beads are explicitly considered in the path integral and in the transition region, in the coarse grained region they are neglected by the fact that the forces acting on them are not calculated and only the dynamical evolution of the coarse-grained sphere is considered.
\bibitem{note2}
 In fact for $T_{1}$ the radial distribution function of the molecular center of mass of the path integral simulation (not shown here) is very close to that of the classical simulation. This means that the coarse-grained potential is very similar to that of the classical case and that the rings are rather stiff and thus the molecular representation in terms of path integral is very close to the atomistic one. We have looked also at the comparison between a full atomistic simulation and a full path integral simulation, at the two temperatures, for the intramolecular bond distribution. For $T_{1}$ we have a broader distribution of the path integral simulation compared to the classical one, however the peak is located more or less at the same point, thus the path integral approach and the classical one will still be relatively close. For $T_{2}$ the distribution is much broader and the peak is shifted and thus represents a test for the validity of the AdResS procedure in this context.
\bibitem{note3}
This in general is acceptable but not highly accurate for realistic applications; anyway we have done extended tests using liquid systems of simple spheres employing $n=20$ and $n=30$ and we obtain basically the same results as $n=10$. Note that the arbitrary choice of $T_{1}$, $T_{2}$ and $n$ is possible because we study a toy model without a specific physical meaning; in real systems $k$ is defined by the physics of the system.
\end{thebibliography}
\end{document}